\documentclass[showpacs,amsmath,amssymb,floatfix,prl,twocolumn]{revtex4}
\usepackage{graphicx,amsmath}

\begin{document} 

\title {Poincar\'e recurrences of DNA sequence}

\author{ K.M.Frahm }
\affiliation{\mbox{Laboratoire de Physique Th\'eorique du CNRS, IRSAMC, 
Universit\'e de Toulouse, UPS, F-31062 Toulouse, France}}
\author{D.L.Shepelyansky}
\affiliation{\mbox{Laboratoire de Physique Th\'eorique du CNRS, IRSAMC, 
Universit\'e de Toulouse, UPS, F-31062 Toulouse, France}}

\date{September 2, 2011}

%

\begin{abstract}
We analyze the statistical properties of Poincar\'e
recurrences of Homo sapiens, mammalian and other 
DNA sequences taken from Ensembl Genome data base
with up to fifteen billions base pairs.
We show that the probability of Poincar\'e recurrences
decays in an algebraic way
with the Poincar\'e exponent $\beta \approx 4$
even if oscillatory dependence is well pronounced.
The correlations between recurrences
decay with an exponent $\nu \approx 0.6$
that leads to an anomalous super-diffusive
walk. However,  for Homo sapiens sequences,
with the largest available statistics,
the diffusion coefficient converges to a finite value
on distances larger than million base pairs.
We argue that the approach based on Poncar\'e recurrences 
determines new proximity features between different species
and shed a new light on their evolution history.
\end{abstract}

\maketitle

The Poincar\'e recurrence theorem of 1890 \cite{poincare} states that 
after a certain time
a dynamical Hamiltonian trajectory in a bounded phase space always
returns to a close vicinity of an initial state.
Even if recurrences definitely take place 
the question about their properties,
or more exactly what are the statistics of Poincar\'e recurrences,
and what are their correlation properties,
still remain an unsolved problem for 
systems of dynamical chaos even after 
an impressive development of the theory of dynamical complexity
\cite{arnold,sinai,lichtenberg}.
The two limiting case of periodic and fully chaotic motion
are well understood: in the first case the recurrences are periodic
while in the latter case the probability of recurrences 
$P(t)$ with time being larger than $t$ drops exponentially 
at $t \rightarrow \infty$ \cite{arnold,sinai,lichtenberg}.
Thus, the latter case is  similar to a coin 
flipping, where a probability to stay on the same side 
after more than $t$ flips decays at $2^{-t}$. However,
in generic Hamiltonian systems
the probability $P(t)$
decays algebraically with $t$, as $P(t) \sim 1/t^\beta$, 
due to long trappings in a vicinity of stability islands 
showing the Poincar\'e exponent $\beta \approx 1.5$
\cite{chirikov1984,ott,chirikov1999,kantz,ketzmerick,dls2010}.
A detailed theoretical explication of this slow algebraic decay
is still lacking. Usually, the consecutive recurrences
in dynamical systems are not correlated
since a trajectory passes across domains of 
chaotic component.

The  Poincar\'e recurrences represent a powerful tool
for analysis of statistical properties of 
symbolic trajectories of various types \cite{arnold,sinai,lichtenberg}.
Surprisingly, this powerful tool of dynamical systems
has not been applied for 
detailed statistical studies of DNA sequence
which also can be viewed as a symbolic trajectory.
There have been only a few earlier attempts going in this direction
including researchers in dynamical systems \cite{turchetti}
and bioinformatics \cite{nair,ferreira1,ferreira2}.
However, in \cite{turchetti} only 
short recurrence times with $t \leq 4$
have been considered and it was concluded that the
probability of recurrences decays exponentially.
The studies in bioinformatics were not aware
about the concept of Poincar\'e recurrences but 
their approach had certain links with them
aiming to use digital signal representations of genomic data
\cite{nair}. The relative frequency analysis
applied in \cite{ferreira1,ferreira2}
has certain similarities with the Poincar\'e recurrences
approach but the distance times
still remain very short with $t \leq 20$ in \cite{ferreira1}
and $t \leq 100$ in \cite{ferreira2}. 
No detailed comparative analysis with exponential decay of
Poincar\'e recurrences of
random sequences  
or algebraic decay was presented there.

In this work,
we apply  the powerful approach of Poincar\'e 
recurrences to available mammalian DNA sequences taken from
the publicly available database
\cite{genbank}. The comparison with random data
sequences and the known results for 
dynamical maps \cite{chirikov1984,ott,chirikov1999,kantz,ketzmerick,dls2010}
allowed us to establish new interesting features 
for the Poincar\'e recurrences of DNA sequence.
Our approach allowed to analyze the recurrences with
time $t$ being by 5 to 6 orders of magnitude larger than
those reached in \cite{turchetti,nair,ferreira1,ferreira2}.
For Homo sapiens (HS) database we performed statistical analysis 
for $1.5 \cdot 10^{10}$ base pairs (bp). 
This amount of statistical data
is by $4 - 5$ orders of magnitude larger
compared to the previous studies
of anomalous diffusion performed  in \cite{peng,kaneko,voss}
for DNA sequences. Using this large statistics
we find  that the DNA Poincar\'e recurrences
are characterized by an algebraic decay
with $\beta \approx 4$ for
HS database. 
For such a value of the Poincar\'e exponent $\beta$, 
the uncorrelated recurrences
should lead to a usual diffusive random 
walk with a linear growth of the corresponding second moment 
$\sigma \sim D t$ \cite{ott,chirikov1999}, with an effective
time $t$ given by the sequence length $L$ 
measured in number of bp.
At the same time the early studies 
for random walk in DNA sequences \cite{peng,kaneko,voss},
with the total length of $t < 10^6$ bp,
established that such a walk belongs to the Levy type walks, with
an anomalous superdiffusive growth of the second moment
$\sigma \sim t^{1+\mu}$ and a growing diffusion  coefficient
$D(t) = \sigma/t \sim t^\mu$ with $\mu > 0$.
Our studies show that this apparent contradiction
is resolved by the presence of long range correlations $C_P(t)$
between the Poincar\'e recurrences in DNA 
that makes them different compared to dynamical chaos
systems where such correlations are usually absent 
\cite{chirikov1984,ott,chirikov1999,kantz,ketzmerick,dls2010}.
We show that $C_P(t)$ is characterized by a 
global algebraic decay with an exponent
$\nu \approx 0.6$. Such a slow decay leads to an 
anomalous super-diffusion on scales of
$t < 10^6$ bp with the exponent $\mu$ being 
in agreement with the previous studies 
\cite{peng,kaneko,voss}. However, for 
$t>10^6$ bp the diffusion coefficient
$D(t)$ for HS becomes finite due to
cancellations of odd and even correlation
terms which show a global algebraic decay with an exponent
$\nu \approx 0.6$. We argue that the obtained results
for the statistics of Poincar\'e recurrences of DNA sequence
open new possibilities for the genome evolution analysis.

\begin{figure}[h]
\begin{center}
\includegraphics[width=0.48\textwidth]{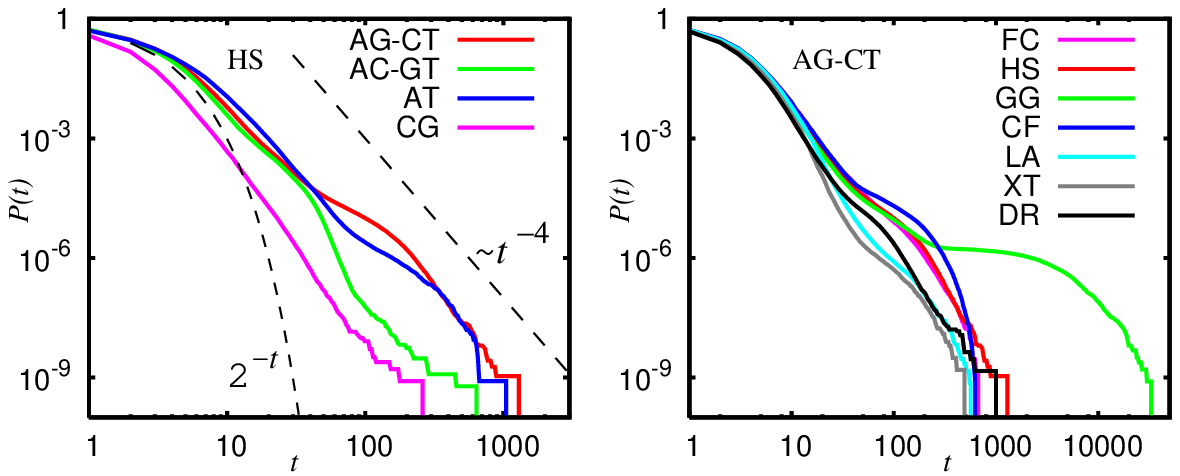}
\includegraphics[width=0.48\textwidth]{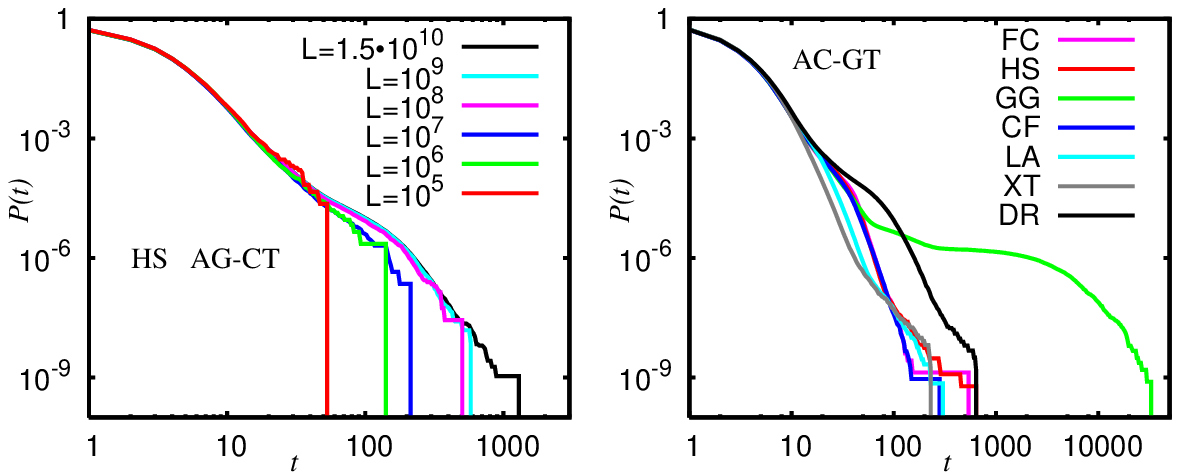}
\end{center}
\vglue -0.2cm
\caption{Statistics of Poincar\'e recurrences $P(t)$ for 
DNA sequences. {\em Top left panel}: DNA data of Homo sapiens (HS) 
for Poincar\'e recurrences of domains AG-CT, AC-GT, AT 
and CG (see text). 
The lower dashed curve shows the exponential behavior $P(t)=2^{-t}$ 
valid for {\em random} sequences, the upper dashed line shows the 
average power law $P(t) \sim t^{-4}$ for comparison. 
{\em Top right panel}: AG-CT data for 
DNA sequences of the species: Felis catus 
(FC, Cat), Homo sapiens (HS, Human), 
Gorilla gorilla (GG, Gorilla), Canis familiaris (CF, Dog), 
Loxodonta africana (LA, Elephant), Xenopus tropicalis 
(XT, African Clawed Frogs) and Danio rerio (DR, Zebrafish).
{\em Bottom left panel}: 
Convergence of the statistics of AG-CT Poincar\'e recurrences $P(t)$
for Homo sapiens as the length $L$ of the considered DNA sequence increases 
from $L=10^5$ to $L=1.5 \cdot 10^{10}$. 
{\em Bottom right panel}:  AC-GT
data sets for the same species as in the top right panel.
\label{fig1}}
\end{figure}

To study the statistics of Poincar\'e recurrence of 
mammalian DNA sequences we use the enormous database \cite{genbank}
considering a DNA sequence as a very long trajectory in the space of four
nucleobases A, G, C, T. Similar to
\cite{peng}, a walk along the DNA sequence length,
marked as an effective time $t$, is described by
a discrete variable $u(t)$ which takes values
``$+$'' for A, G of purine domain and ``$-$'' for
C, T of pyrimidine domain (AG-CT). 
The differential distribution of Poincar\'e recurrences
$p_1(t)$ is given by a relative number of segments
of fixed sign of length $t$ while the integrated
distribution $P(t)$ gives the relative number of recurrences
with times larger than $t$. The probabilities of
domains AG and CT are close to $0.5$ for HS
and mammalian sequences.
Thus the recurrences for both domains are 
very close to each other so that we show 
one average distribution $P(t)$ for AG-CT
corresponding to recurrences or crossings of line $u=0$.
A similar situation takes place for 
AC and  GT domains  so that
we show for them one average distribution $P(t)$ for AC-GT.
For domains AT and CG the probabilities are approximately
$0.6$ and $0.4$ and here we show separately recurrence probability
$P(t)$ for AT and CG domains. 
For Poincar\'e recurrences $P(t)$ of HS sequences
these four cases are shown in Fig.~\ref{fig1} (left panel).
In average we find an algebraic decay $P(t) \sim 1/t^\beta$
with $\beta \approx 4$. A formal fit for AG-CT data at $t>10$ gives
$\beta = 3.68 \pm 0.02$ but there are visible large scale
oscillations with a certain similarity to
those seeing in dynamical maps \cite{chirikov1984,chirikov1999,ketzmerick}.
The dependence $P(t)=2^{-t}$ for a random sequence 
describes AG-CT and AC-GT data only on short times $t < 5$
while for larger times algebraic behavior becomes dominant.
We note that $P(t)$ is a positively defined quantity
and thus it is statistically very stable: 
the sequences of size $L$ well reproduce 
the initial part of $P(t)$ almost up to values $\sim 1/L$
as it is shown in Fig.~\ref{fig1} 
(bottom left panel), where $L$ varies in a large interval
of $10^5 \leq L \leq 1.5 \cdot 10^{10}$ bp. 

The comparison of statistics of Poincar\'e recurrences
for HS, mammalian and two other species are shown in Fig.~\ref{fig1}
for AG-CT case (similar average behavior is found for AC-GT data).
The total sequence lengths $L$ for other species are
by a factor 3 shorter compared to HS case. 
Up to $t \approx 20$ all considered species show
the same decay of $P(t)$ but at larger value of $t$
there is a separation of curves so that
each species is characterized by
its own statistics $P(t)$. In average all species
show an algebraic decay with $\beta \approx 4$
even if there is a strong oscillation
with a flat region of $P(t)$ for GG sequence 
(AC-GT data from Fig.~\ref{fig1} show a very similar behavior in this case).
It is interesting to note that
the curves of Poincar\'e recurrences are 
very close for HS and GG sequences up to $t \approx 200$
and for HS and FC sequences up to maximal $t \approx 10^3$.
However, for AC-GT data set the curves for these sequences become
different for $t > 20$ (Fig.~\ref{fig1}). 
\begin{figure}[h]
\begin{center}
\includegraphics[width=0.48\textwidth]{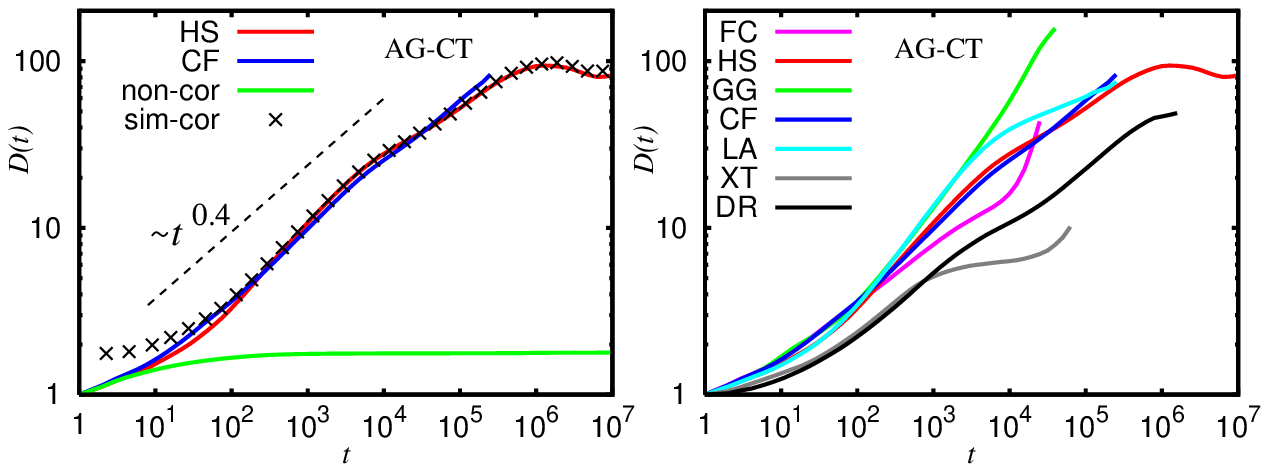}
\includegraphics[width=0.48\textwidth]{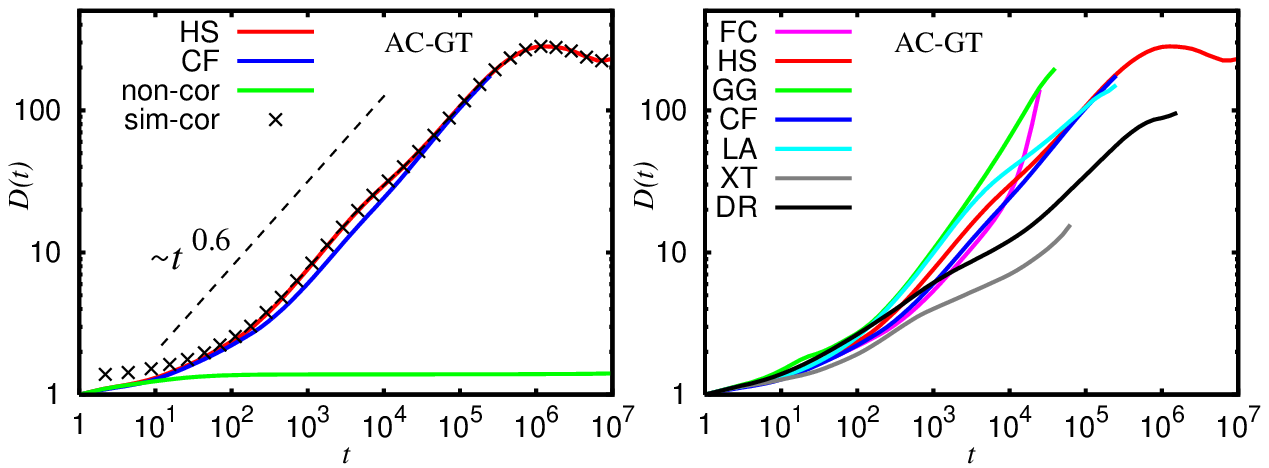}
\end{center}
\vglue -0.2cm
\caption{{\em Top left panel}: Diffusion coefficient
$D(t)=\langle \Delta y^2(t)\rangle/t$ for AG-CT data sets of
DNA sequences of HS
and CF. The lower green curve (non-cor) 
is the diffusion coefficient obtained for a model with individual 
recurrences being distributed as the Poincar\'e 
recurrences of HS in Fig.~\ref{fig1} 
but {\em assuming} that subsequent 
Poincar\'e recurrences are {\em not correlated}. 
The black crosses (sim-cor) represent the diffusion coefficient obtained 
from (\ref{eq2}) using the 
Poincar\'e recurrence correlation function
$C_P(n)$ for HS 
(see text and Fig.~\ref{fig4} below). 
The dashed line shows a power law $D \sim t^{0.4}$. 
{\em Top right panel}: 
Diffusion coefficient $D(t)$ for AG-CT data sets
of the same species as in the right panel 
of Fig.~\ref{fig1}.
 {\em Bottom panels:} Diffusion coefficient $D(t)$
for AC-GT data sets for the same cases as in top panels;
the dashed line in the left panel 
represents a power law dependence $D(t) \sim t^{0.6}$.
\label{fig2}} 
\end{figure}

It is important to understand how the statistics of Poincar\'e recurrences
is related to the anomalous 
super-diffusive walk discussed in \cite{peng,kaneko,voss}.
The walk is described by a displacement variable
$y(t)=\sum_{\tau=1}^t u(\tau)$ whose growth can be characterized by
a diffusion coefficient defined as 
$D(t)=\sigma / t$ with the second moment
$\sigma=\langle \Delta y(t)^2\rangle$,
$\Delta y(t)=y(t+t_0)-y(t_0)-\langle y(t+t_0)-y(t_0)\rangle$ and 
the average $\langle\cdots\rangle$ is done with respect to the initial 
position (or ``time'') $t_0$. In case of a standard diffusive process
the diffusion coefficient $D$
converges to a finite value at large times. 
However, the results of \cite{peng} give an algebraic super-diffusive
growth $D(t)\sim t^{\mu}$ with the exponent $\mu  \approx 0.34$
for HS sequence of length $L \sim 10^5$ and $t \leq 10^3$.
Our results are obtained on a significantly larger scale of $t$
being by 4 orders of magnitude larger compared to those
reached in \cite{peng,kaneko,voss}.
Our  results for diffusion $D(t)$ are shown in Fig.~\ref{fig2}. 
For HS sequence 
we have large statistics and large exact segments
without non-determined bp marked as $N$ in database \cite{genbank}.
We find $\mu \approx 0.4$ for the range $10 < t < 10^6$
(fit gives $\mu = 0.349 \pm 0.001$) in a satisfactory agreement
with previous studies \cite{peng,kaneko,voss}.
Other species also show an algebraic growth of $D(t)$
with similar values of $\mu$ (Fig.~\ref{fig2}).
For AC-GT data we also find a similar behavior
with $\mu \approx 0.6$ for HS sequence (Fig.~\ref{fig2}). 
However, for HS sequence with most exact and long
data set we find a saturation of $D(t)$
for large times $10^6 \leq t \leq 10^7$.

The diffusion coefficient is related to the correlation function 
$c(t)=\langle u(t+t_0) u(t_0)\rangle$ as 
$D(t)=(1/t)\sum_{l=1}^t\sum_{j=-l+1}^{l-1} c(j)$
and hence a divergence of $D$ implies
a slow correlation decay $c(t) \sim t^{\mu -1}$
if $c(t)$ is monotonic. 
On the other hand this correlation function can also be expressed as
\begin{equation}
\label{eq0}
c(t)=\sum_{n=1}^\infty (-1)^{n-1}
\!\!\!\!\!\!\!\!\!
\sum_{
t_1+\ldots+t_n>t 
}^\infty
\!\!\!\!\!\!
(t_1+\ldots+t_n-t)\,p_n(t_1,\,\ldots,\,t_n)
\end{equation}
where $p_n(t_1,\,\ldots,\,t_n)$ is the joint distribution of 
$n$ subsequent Poincar\'e recurrence times $t_1,\,\ldots,\,t_n$. 
In this sum each term represents the case where $n$ subsequent recurrences are 
needed to cover the interval $0,\,1,\,\ldots,\,t$ and the prefactor 
$t_1+\ldots+t_n-t$ accounts for the number of different initial positions 
of the first recurrence to allow this. If we {\em assume} that 
subsequent Poincar\'e recurrences are {\em not correlated}, i.~e.:
$p_n(t_1,\,\ldots,\,t_n)=p_1(t_1)\cdot\ldots\cdot p_1(t_n)$, 
and that $P(t_1)$ obeys the power law $P(t_1)\sim t_1^{-\beta}$, i.~e.: 
$p_1(t_1)=P(t_1)-P(t_1+1)\sim t_1^{-\beta-1}$ we find that in 
the above expression the first term for $n=1$ dominates the limit 
$t\to\infty$ and we find that 
$c(t)\approx \sum_{t_1=t+1}^\infty P(t_1)\sim t P(t)\sim t^{1-\beta}$. 
We mention that this result was previously also obtained for 
chaotic Hamiltonian dynamics \cite{ott,chirikov1999}.
Therefore we should have a good convergence of $D$
with $\beta \approx 4$. 
However, this relation is obtained
for the case of {\em uncorrelated} Poincar\'e recurrences
that may not be the case for DNA sequences. Indeed, 
if we generate uncorrelated  recurrences with the distribution
$P(t)$ being the same as in Fig.~\ref{fig1} for AG-CT sequence of HS
and compute with them the diffusion coefficient then we find
a clear saturation of $D(t)$ at a finite value $D =1.77$
(green curve in Fig.~\ref{fig2}, left panel),
being significantly smaller then the actual data of $D(t) \sim 100$. 
\begin{figure}[h]
\begin{center}
\includegraphics[width=0.48\textwidth]{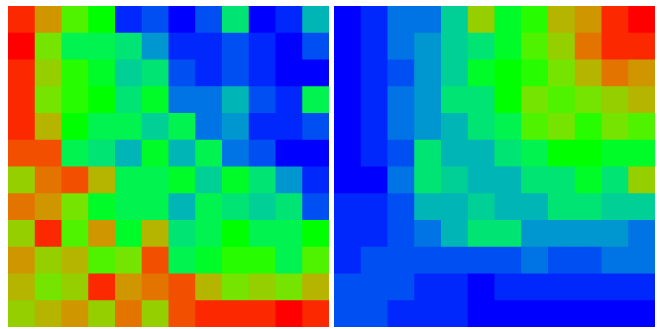}
\end{center}
\vglue -0.2cm
\caption{{\em Left panel}: Density plot of the normalized 
two point correlator $\tilde{p}_2(t_1,t_2)$ of two 
subsequent Poincar\'e recurrences $t_1$ and $t_2$ 
for AG-CT data sets of HS. The shown range 
$1\le t_1,t_2\le 12$ represents $99.4$\% of probability. 
Red (green, blue) color represents maximal (zero, minimal) values, 
horizontal/vertical axes show $t_1$ and $t_2$.
{\em Right panel}: Normalized two point correlator 
$\tilde{p}_2(t_1,t_3)$
of $t_1$ and $t_3$ for three subsequent Poincar\'e recurrences 
$t_1$, $t_2$, $t_3$ with $t_1$ and $t_3$ on the axes. 
\label{fig3}}
\end{figure}

To visualize the correlations between Poincar\'e
recurrences we also compute the joint probability 
$p_2(t_1,t_2)$ of two subsequent  Poincar\'e recurrences $t_1$ and $t_2$
for HS sequence of Fig.~\ref{fig1}. The normalized two point 
correlator is $\tilde{p}_2(t_1,t_2)=p_2(t_1,t_2)/[p_1(t_1)\,p_1(t_2)]-1$, 
where $p_1(t_1)=P(t_{1})-P(t_{1}+1)$ is the probability 
of one individual recurrence of length $t_{1}$. 
Its dependence on $t_1$, $t_2$ is shown in Fig.~\ref{fig3}.
The  correlator is maximal for $t_1=1$ (i.e. below the average 
recurrence time $\langle t_1 \rangle=2.27$) 
and $t_2\ge 8$ (i.e. above average) or vice-versa 
thus indicating anti-correlations between $t_1$ and $t_2$.
In the right panel of Fig.~\ref{fig3} we show 
the normalized two point correlator  $\tilde{p}_2(t_1,t_3)$
for $t_1$ and $t_3$ taken from three subsequent Poincar\'e recurrence times 
$t_1$, $t_2$, $t_3$. In this case $t_1$ and $t_3$ are correlated, i.e. if 
$t_1$ is above average it is more likely that $t_3$ is also above average.

Thus, in a sequence of Poincar\'e recurrences $t_1,\,t_2,\,t_3,\,\ldots$ 
the odd elements represent steps of length $t_1,\,t_3,\,\ldots$ of one sign 
of $u(t)$ and the even elements represent 
steps of length $t_2,\,t_4,\,\ldots$ of the other sign.
The anti-correlations between $t_1$ and $t_2$ or $t_2$ and $t_3$ 
as well as the correlations between $t_1$ and $t_3$ indicate that once 
a preferential direction is chosen it is more likely 
for it to be enhanced thus 
explaining the diffusion enhancement 
compared to the uncorrelated Poincar\'e recurrences 
which give a finite coefficient $D \approx 1.7$. 
To work out this point on a more quantitative level we 
consider the displacement after $n$ Poincar\'e recurrences at 
time $t=t_1+\ldots+t_n\approx n\langle t_1\rangle$.
We can write for it
\begin{equation}
\label{eq1}
y(t_1+\ldots+t_n)=(-1)^s\sum_{l=1}^n (-1)^{l-1} t_l \;\; ,
\end{equation}
where $(-1)^s$ is the sign of the first segment associated to $t_1$. 
For $n\gg 1$ this leads to 
\begin{equation}
\label{eq2}
D(n\langle t_1\rangle)=\frac{1}{n\langle t_1\rangle}\,
\sum_{l=1}^n\Bigl(C_P(0)+2\sum_{j=1}^{l-1} (-1)^j\,C_P(j)\Bigr) \; ,
\end{equation}
where $C_P(j)=\langle t_1\,t_{1+j}\rangle-\langle t_1\rangle^2$ is 
the Poincar\'e recurrence correlation function
and the average is done over all recurrences \cite{footnote_correl}.
We note that the above model of uncorrelated Poincar\'e 
recurrences  corresponds to $C_P(j)=0$ for $j>0$. In this 
case Eq.(\ref{eq2}) gives 
$D=C_P(0)/\langle t_1\rangle = 4.01/2.27 = 1.77$
in a perfect agreement 
with the data of Fig.~\ref{fig2}.

\begin{figure}[h]
\begin{center}
\includegraphics[width=0.48\textwidth]{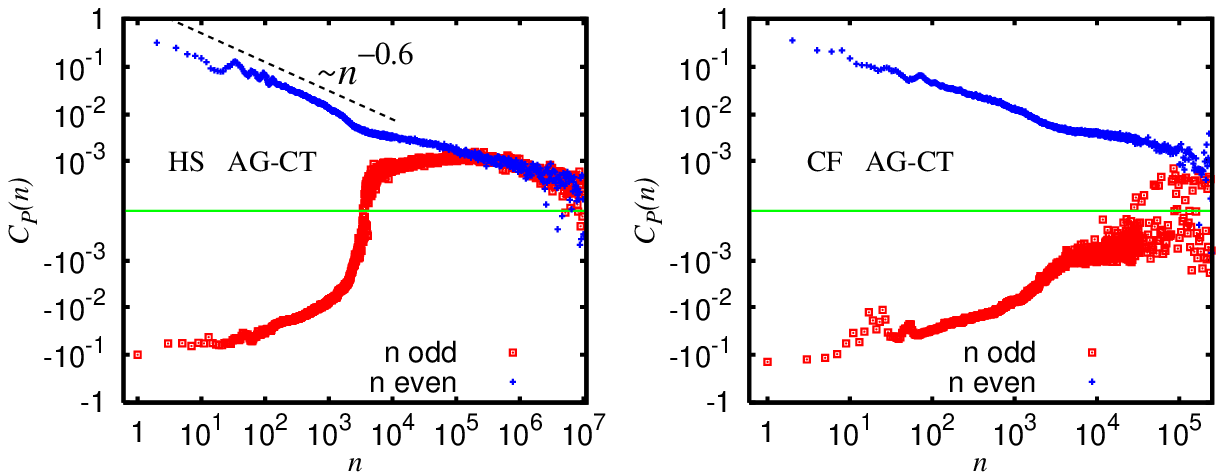}
\includegraphics[width=0.48\textwidth]{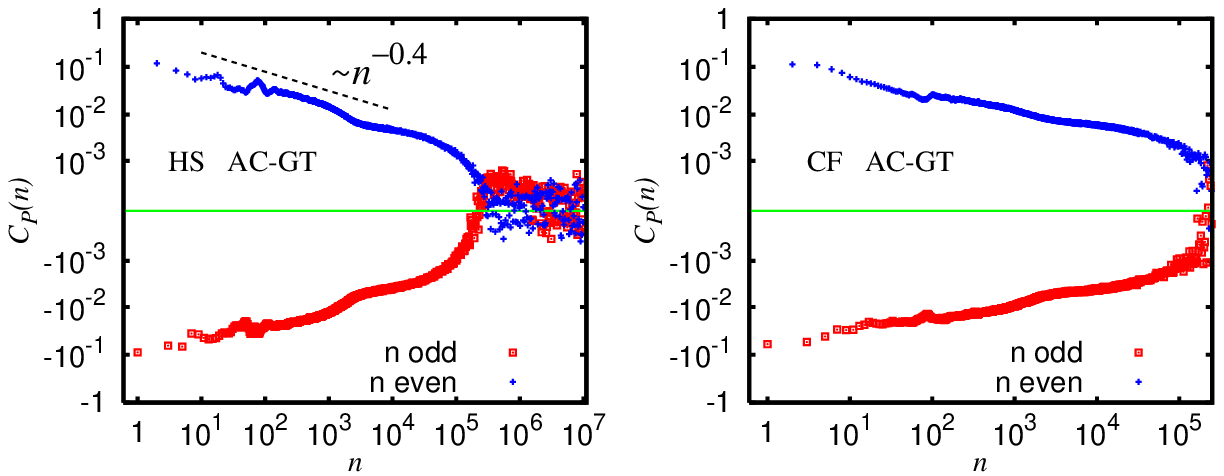}
\end{center}
\vglue -0.2cm
\caption{{\em Top panels:} 
Poincar\'e recurrence correlation function 
$C_P(n)=\langle t_1\,t_{n+1}\rangle - \langle t_1\rangle^2$ of $t_1$ and 
$t_{n+1}$ in a sequence of subsequent Poincar\'e recurrences $t_j$ 
for HS (left panel) and CF (right panel) sequences 
for AG-CT data sets. 
Blue crosses correspond to even $n$ and red squares to odd $n$. The 
dashed line shows a power law $C_P(n) \sim n^{-0.6}$. 
For clarity positive and negative values of $C_P(n)$ are 
shown on two separate logarithmic scales which are put together at 
$C_P=\pm 10^{-4}$ shown by the green line.
{\em Bottom panels:} same as in top panels but 
for AC-GT data sets 
for HS (left panel) and 
CF (right panel);
the dashed line shows the dependence
$C_P(n) \sim n^{-0.4}$.
\label{fig4}}
\end{figure}

\begin{figure}[h]
\begin{center}
\includegraphics[width=0.48\textwidth]{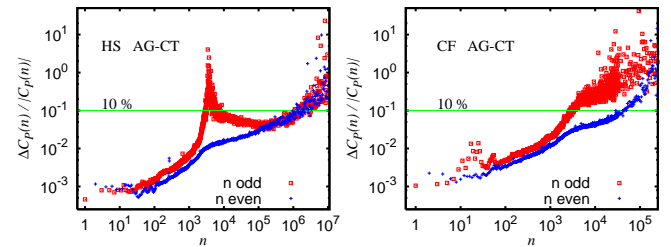}
\end{center}
\vglue -0.2cm
\caption{Relative statistical error $\Delta C_P(n)/|C_P(n)|$ 
of the Poincar\'e recurrence correlation function 
for HS (left panel) and CF (right panel) sequences 
for AG-CT data sets shown in top panels of Fig.~\ref{fig4}. 
Blue crosses correspond to even $n$ and red squares to odd $n$. 
The green line indicates the value of 10 \%. 
\label{fig5}}
\end{figure}

The Poincar\'e recurrence correlation function 
$C_P(n)$ is computed from DNA sequence data and 
its dependence on recurrence index/number $n \approx t/\langle t_1 \rangle$ 
is shown 
in Fig.~\ref{fig4} for AG-CT data sets of HS and CF.
For HS data this correlation function has alternate signs
for odd and even $n$ up to $n \approx 3 \cdot 10^3$. 
For larger $n$ values these terms have
the same sign and moreover these terms become approximately equal
for $n > 10^5$. This leads to cancellation
of odd and even terms in (\ref{eq2}) and saturation
of the growth of diffusion coefficient at $t > 10^6$
as it is clearly seen in Fig.~\ref{fig2}.
Such a saturation of $D(t)$ takes place 
in spite of a rather slow algebraic decay of correlation
$C_P(n) \sim n^{-\nu}$ with $\nu \approx 0.6$
(for even terms an error weighted fit gives $\nu = 0.575 \pm 0.003$
at $10 \leq n \leq 3 \cdot 10^6$ and for odd terms $\nu = 0.479 \pm 0.005$
at $10 \leq n \leq 10^3$).  
From the found 
correlation function $C_P(n)$  
we can determine the dependence $D(t)$ using (\ref{eq2})
that gives a good agreement with
the data obtained by a direct computation of 
$D(t)$ as it is shown in Fig.~\ref{fig2}
(deviations at $ t  <10$ are due to 
an approximate validity of the relation
$t=t_1+..+t_n \approx n \langle t_1 \rangle$ at small $t$). 
We note that the relation between exponents
$\mu = 1-\nu$, corresponding to a
simple estimate $D \sim t\,|C_P(t)|$, remains
valid in absence of odd/even terms cancellation
at $t < 10^6$. For CF data set we find approximately
the same algebraic decay with $\nu \approx 0.6$
(Fig.~\ref{fig4}, right panel). In this case the 
total number of recurrences $N_r$
is statistically smaller compared to HS case
and in addition undetermined letters $N$ of bp
are broadly scattered over the
sequence. Due to that here we do not find
large number $N_r(n)$
of recurrence times at large $n$
that force us to stop 
at $n<2.5 \cdot 10^5$ where
a saturation of $D(t)$ growth is not visible
(for HS case we have $N_r(n) \approx 5 \cdot 10^9$
recurrences at $n=10^6$ but many of them are
correlated and the statistical error of
$C_P(n)$ is about 5\% here while for
smaller $n$ it becomes smaller 
than the symbol size in Fig.~\ref{fig4}). 
For AC-GT data sets,
shown in Fig.~\ref{fig4}, we find an
algebraic decay with exponent $\nu \approx 0.4$
corresponding to the value $\mu \approx 0.6$
from corresponding Fig.~\ref{fig2}. 
The convergence of odd/even terms of $C_F(n)$
for HS case takes place at $n > 10^5$
leading to saturation of diffusion rate
at $t>10^6$ also visible for AC-GT data
(Fig.~\ref{fig2}). For CF data
we have lower statistics for large $n$ and $t$
and saturation of $D(t)$ remains invisible.

The analysis of statistical accuracy of the 
computation of correlation function $C_P(n)$
is presented in Fig.~\ref{fig5}.
Here we show the variation 
of relative statistical error $\Delta C_P(n)/|C_P(n)|$
in the value of  $C_P(n)$ as a function
of $n$. This error increases
from a level of $10^{-3}$ at $n <100$
up to $0.1$ at $n \approx 2 \cdot 10^6$
for HS and at $n \approx 10^4$ for CF
(a strong increase of error at 
$n\approx 3 \cdot 10^3$ for HS
is related to a sign change of 
$C(n)$). The relative error increases with
$n$ since at large $n$ we have smaller
number of recurrences $N_r$ contributing in the
computation of $C(n)$. For the HS case the number
of non-determined  $N$ letters allows to
have a significantly larger number of recurrences 
$N_r$ compared to
the CF case and due to this we obtain
statistically good values of $C(n)$ at
significantly larger values of $n$.

Let us give now the formal fit parameter values
for the dependencies discussed above.
The fit of Poincar\'e recurrences 
for the data of Fig.~\ref{fig1} at $t>10$
gives the Poincar\'e exponent 
$\beta= 3.68 \pm 0.02$ (AG-CT),
$3.65 \pm 0.04$ (AC-GT), $3.75 \pm 0.03$ (AT),
$4.04 \pm 0.05$ (CG). The fit of $D(t) \sim t^\mu$
for AG-CT data of HS in Fig.~\ref{fig1}
gives $\mu = 0.3486 \pm 0.0008$ for the range
$10 \leq t \leq 10^6$ but there are
two intervals with distinct values
$\mu =0.5010 \pm 0.0003$ for
$10 \leq t \leq 3 \cdot 10^3$ and
$\mu =0.2859 \pm 0.0003$ for
$3 \cdot 10^3 \leq t \leq 10^6$
so that we give in the text the average $\mu \approx 0.4$.
For AC-GT data of HS the whole range
of $D(t)$ is well characterized by
a fit exponent  $\mu = 0.5553 \pm 0.0004$
for $100 \leq t \leq 10^6$ (see Fig.~\ref{fig2}).
Furthermore for AC-GT data the correlation function behaves also as 
$C_P(n) \sim n^{-\nu}$ where for HS 
the exponent obtained from an error weighted fit is 
$\nu = 0.367 \pm 0.004$ for even terms and $\nu = 0.320 \pm 0.004$ for odd 
terms, both at $10 \leq n \leq 10^4$.
Even if formal statistical errors are quite small
we should note that there are
rather pronounced oscillations
and due to that reason we
give in the above discussions only
approximate values of the exponents.

The presented results determine the 
statistics of Poincar\'e recurrences
of DNA sequences and link their
properties to the statistics of sequence walks
studied previously \cite{peng,kaneko,voss}.
The anomalous diffusion of walks
is related to enormously long
correlations between far away recurrences.
For most detailed HS sequences
the diffusion coefficient of these walks
becomes finite due to cancellations
of slow decaying correlations.
For other species larger statistical samples
are required to see if the diffusion
coefficient saturation is present.
The Poincar\'e recurrences $P(t)$
are statistically very stable 
and show clear difference
between various species. 
The statistical analysis of human and mammalian 
DNA sequences is now an active research field
with links to genome evolution
(see e.g. \cite{gibbs,galtier,trifonov})
and the approach based on Poincar\'e recurrences
should bring here new useful insights.

The obtained properties of Poincar\'e recurrences
can be used for verification of various theories of genome evolution
(see e.g. \cite{gibbs,galtier,trifonov,nei,cooper}).
Such theories should reproduce well the main 
statistical features of
Poincar\'e recurrences described here.
Indeed, the data of Fig.~\ref{fig1} show that for
$t < 5$ the recurrences for all analyzed species
behave like a random sequence of coin flipping.
Thus the genome evolution 
generates random uncorrelated 
short range recurrences. However,
for the range $5 \leq t \leq 20$
we have  a beginning of algebraic decay of $P(t)$
but still all the species follow practically the same 
curve. 
This indicates an existence of 
a common period of initial 
evolution history.
For $t>20$ we observe a strong divergence
of Poincar\'e curves of different species.
Surprisingly the curves of HS and FC
(as well as LA and XT)
remain very close to each other up to 
largest recurrences with $t \approx 400$
for AG-CT data sets. At the same time
for AC-GT data sets a close proximity
of recurrences is observed for
HS, LA and XT (as well as for FC and CF)
up to largest values $t \approx 300$.
This show various aspects of 
proximity between species which
should be investigated in further studies.
We hope that the new tool of Poincar\'e recurrences
will allow to analyze the proximity between species
under a new angle  lightening new sides of 
life evolution.


\begin{thebibliography}{99}
\bibitem{poincare} H.~Poincar\'e, 
        Acta Math. {\bf 13}, 1 (1890).
\bibitem{arnold} V.I.Arnold and A.Avez, {\it Ergodic problems of
        classical mechanics}, Benjamin, Paris (1968).
\bibitem{sinai} I.P.~Cornfeld, S.V.~Fomin and Y.G.~Sinai,
        {\it Erodic theory}, Springer, N.Y. (1982).
\bibitem{lichtenberg} A.J.~Lichtenberg and M.A.~Lieberman,
        {\it Regular and chaotic dynamics}, Springer, Berlin (1992).
\bibitem{chirikov1984} B.V.~Chirikov and D.L.~Shepelyansky,
        Physica D {\bf 13}, 395 (1984).
\bibitem{ott} J.D.~Meiss and E.~Ott,
        Phys. Rev. Lett. {\bf 55}, 2741 (1985).
\bibitem{chirikov1999} B.V.~Chirikov and D.L.~Shepelyansky,
        Phys. Rev. Lett. {\bf 82},  528 (1999); 
        {\bf ibid.} {\bf 89}, 239402 (2002).
\bibitem{kantz} E.G.~Altman and H.~Kantz,
        Europhys. Lett. {\bf 78}, 10008 (2007).
\bibitem{ketzmerick} G.~Cristadoro and R.~Ketzmerick, 
        Phys. Rev. Lett. {\bf 100},  184101 (2008).
\bibitem{dls2010} D.L.~Shepelyansky, 
        Phys. Rev. E {\bf 82}, 055202(R) (2010).
\bibitem{turchetti} L.~Rossi and G.~Turchetti,
         Physica A {\bf 338}, 267 (2004).
\bibitem{nair} A.S.S.Nair and T.~Mahalakshmi,
         Proceedings of IEEE
         Genomic Signal Processing. Bucharest, Romania (2005)
\bibitem{ferreira1} V.~Afreixo, C.A.C.~Bastos, A.J.~Pinho, S.P.~Garcia
         and P.J.S.G.~Ferreira, 
          Bioinformatics {\bf 25}, 3064 (2009)
\bibitem{ferreira2} C.A.C.~Bastos, V.~Afreixo, A.J.Pinho, S.P.~Garcia,
         J.M.O.S.~Rodrigues and P.J.S.G.~Ferreira,
         Adv. Intel. Soft Comp. (Eds. M.P.Rocha {\it et al.}),  
         Springer, Berlin {\bf 93}, 205 (2011)
\bibitem{genbank} Ensembl Genome data base  http://www.ensembl.org/ and 
                  ftp://ftp.ensembl.org/pub/release-62/genbank/
\bibitem{peng} C.K.~Peng, S.V.~Buldyrev, A.L.~Goldberger,
               S.~Havlin, F.~Sciortino, M.~Simons and H.E.~Stanley, 
                Nature {\bf 356}, 168 (1992). 
\bibitem{kaneko} W.~Li and K.~Kaneko,
               Europhys. Lett. {\bf 17}, 655 (1992).
\bibitem{voss} R.F.~Voss, 
               Phys. Rev. Lett. {\bf 68},  3805 (1992).

\bibitem{footnote_correl} We note that is important to evaluate very 
carefully the Poincar\'e recurrence correlation function as 
$C_P(j)=\langle t_1\,t_{1+j}\rangle-\langle t_1\rangle\,\langle t_{1+j}\rangle$
where the averages $\langle t_1\rangle$ and $\langle t_{1+j}\rangle$ are 
computed as {\em different} quantities using exactly the {\em same} data 
used to compute the average $\langle t_1\,t_{1+j}\rangle$, i.~e. all 
sequences of at least $j+1$ Poincar\'e recurrences $t_1,\,\ldots,\,t_{1+j}$ 
for which the covered DNA sequence is not interrupted by any non-determined
$N$ letter entry. 
In principle the average $\langle t_1\rangle$ can be also  computed for 
a larger data-set (of simply all Poincar\'e recurrences available) but the 
resulting value would be slightly different and not appropriate for the use in 
the correlation function.

\bibitem{gibbs} D.A.~Wheller {\it et al.}, 
              Nature {\bf 452}, 872 (2008).
\bibitem{galtier} J.~Romiguier, V.~Ranwez, E.J.P.~Douzery and
              N.~Gatltier, 
              Genome Res. {\bf 20}, 1001 (2010).
\bibitem{trifonov} Z.M.~Frenkel, T.~Bettecken and E.N.~Trifonov,
              BMC Genomics {\bf 12}, 203 (2011).
\bibitem{nei} M.~Nei, {\it Molecular Evolutionary Genetics},
              Columbia Univ. Press, N.Y. (1987).
\bibitem{cooper} D.N.Cooper, {\it Human gene evolution},
              Elsevier, Amsterdam (1999).

\end{thebibliography}
\end{document}